\documentclass[onecolumn, 12pt, draftclsnofoot, journal]{IEEEtran}

\usepackage{xcolor}
\usepackage{rotating,graphics,psfrag,epsfig}
\usepackage{url}

\begin{document}

\title{Communication and Control Co-Design in 6G: Sequential Decision-Making with LLMs}

\author{Xianfu Chen,~\IEEEmembership{Senior Member,~IEEE}, Celimuge Wu,~\IEEEmembership{Senior Member,~IEEE}, \\Yi Shen, Yusheng Ji,~\IEEEmembership{Fellow,~IEEE}, Tsutomu Yoshinaga,~\IEEEmembership{Member,~IEEE}, \\Qiang Ni,~\IEEEmembership{Senior Member,~IEEE}, Charilaos C. Zarakovitis,~\IEEEmembership{Member,~IEEE}, \\and Honggang Zhang,~\IEEEmembership{Fellow,~IEEE}

%
%
%
%
%
%

}


\maketitle

\begin{abstract}

This article investigates a control system within the context of six-generation wireless networks.
The control performance optimization confronts the technical challenges that arise from the intricate interactions between communication and control sub-systems, asking for a co-design.
Accounting for the system dynamics, we formulate the sequential co-design decision-makings of communication and control over the discrete time horizon as a Markov decision process, for which a practical offline learning framework is proposed.
Our proposed framework integrates large language models into the elements of reinforcement learning.
We present a case study on the age of semantics-aware communication and control co-design to showcase the potentials from our proposed learning framework.
Furthermore, we discuss the open issues remaining to make our proposed offline learning framework feasible for real-world implementations, and highlight the research directions for future explorations.

\end{abstract}


%
\IEEEpeerreviewmaketitle

\section{Introduction}

Wireless networked control systems (NCSs) have been focal in contemporary engineering and industrial applications, owing to the flexibility, scalability and cost-savings \cite{Zhang2020Jan}.
In contrast to the traditional wired NCSs, the design of communication and control policies presents unique challenges due to the stochastic nature of wireless networks.
Wireless networks typically exhibit the spatial and temporal variations in channel quality and radio resource availability, implying that a wireless NCS operates amidst randomness \cite{Ayan2024Apr}.
Consequently, optimizing the stability of a wireless NCS necessitates the efficient radio resource management as well as the reliable control operations.

In a wireless NCS, the communication and control sub-systems are highly interdependent.
On one hand, the communication sub-system facilitates data exchange among the nodes.
The communication performance has a direct impact on the stability of the closed-loop control sub-system \cite{Zheng2023Sep}.
On the other hand, changing the parameters (e.g., the sampling frequency) in the control sub-system leads to data traffic fluctuations over the communication sub-system \cite{Ballotta2022}.
It is evident that independent design of each of the communication and control sub-systems is not optimal.
Accordingly, optimizing a wireless NCS demands a co-design from the perspectives of both communication and control.
Towards this direction, there have been numerous related efforts.
For example, in \cite{Girgis2021Oct}, Girgis et al. applied the Lyapunov drift-plus-penalty framework to optimize the communication and control co-design, the purpose of which is to minimize the network-wide average age of information (AoI) and transmit power.
In \cite{Cao2024}, Cao et al. formulated the stability problem of a wireless NCS as a constrained Markov decision process (MDP) and derived a corresponding goal-oriented deterministic policy.
In \cite{Xu2020}, Xu et al. adopted reinforcement learning (RL) techniques to automatically configure the wireless NCS under the dynamic industrial environment.

Implementing existing communication and control co-design solutions in real wireless NCSs is, however, hindered by the practical and technical obstacles.
Current wireless networks have been primarily designed with a general communication purpose, offering massive connectivity with low latency and high reliability, rather than supporting control applications \cite{Saad2020}.
Consequently, the control functions are not always provisioned.
Moreover, it is impractical to fully and accurately model the intricate interaction between communication and control sub-systems under dynamics, which renders the conventional model-based techniques infeasible for the communication and control co-design.
Leveraging model-free RL to optimize a wireless NCS, there are still the following drawbacks: 1) a large number of intrinsic system features increase the problem size; and 2) the convergence during the online learning process potentially jeopardizes the system stability.
It is anticipated that the six-generation (6G) wireless networks will expedite the convergence of intelligence with communication, computing, control and sensing functions \cite{Letaief2022}.
6G is envisioned as a versatile system that promotes the communication and control co-design.

In the realm of 6G wireless networks, we formulate the sequential co-design decision-makings of communication and control over the infinite time horizon as a model-free RL problem.
To overcome the aforementioned drawbacks of RL techniques, we resort to large language models (LLMs) to aid in the finding of an optimal policy.
The use of LLMs for communication and control co-design facilitates intelligent and efficient management of a wireless NCS in 6G.
LLMs have demonstrated the potentials in unearthing meaningful features concealed in vast amounts of data \cite{Zhao2023Nov}.
This enables the controller (i.e., an RL agent) to understand the wireless NCS from diverse sources of observations, and to use the features to represent the system state, which serves as an input for jointly adjusting communication and control parameters.
With the reasoning capability, LLMs can discern the interrelationships of parameters of communication and control co-design, thereby allowing the controller to abstract an action space.
Concurrently, LLMs decipher the complex patterns behind the received feedbacks, empowering the controller to learn and adapt to the varying requirements of the wireless NCS.
In particular, the offline amalgamation of LLMs and RL is of great importance for a wireless NCS, where the system stability, control reliability and data safety are paramount.

We organize the rest of this article as follows.
In Section \ref{challenges}, we highlight the challenges of communication and control co-design for a wireless NCS in 6G.
In Section \ref{framework}, we discuss in details how LLMs can be integrated into different RL elements, and present a case study to show the potential of our proposed learning framework.
In Section \ref{directions}, we describe the remaining open issues and the research topics that are worth further exploration.
Finally, Section \ref{conclusions} draws the conclusions.

\section{Challenges of Communication and Control Co-Design in 6G Networks}
\label{challenges}

Given the heterogeneity and complexity of a wireless NCS in 6G networks, we have the following specific technical challenges of communication and control co-design.


\subsection{Communication Controls}

In 6G, a wireless network enables the communications among spatially dispersed nodes of a control system with enhanced flexibility and mobility.
For such a wireless NCS, the choice of communication parameters needs to guarantee the stability robustness.
The key parameters of the communication sub-system include transmit power and frequency band.
The challenge, however, lies in the inherent trade-off between latency and reliability, which directly affect the stability of the control sub-system.
Consequently, it becomes essential to devise efficient radio resource management mechanisms that aim to fulfill the low latency and ultra-reliability requirements of a wireless NCS.
The current literature assumes a priori statistical models under stationary environments, failing to adequately capture the control sub-system dynamics.
Moreover, maintaining ultra-reliability while ensuring high communication throughput is an extremely challenging task due to the unpredictable nature of wireless communication conditions.

\subsection{Control Communications}

In the control sub-system, the primary control parameters consist of data sampling and processing as well as node scheduling.
Constrained by the limited radio resources, the wireless communications introduce the latency and the unreliability.
This necessitates the optimization of control parameters to ensure that the control sub-system achieves a certain performance, while being resilient to the impairments caused by the communication sub-system.
However, the majority of existing efforts are built upon the simplistic and restrictive communication models.
For a practical wireless NCS, the communication and control parameters are highly correlated with one another.
That is, the changes in control parameters simultaneously lead to the variations in radio resource consumption, and the vice versa.
It is natural that the neglect of interactions between the communication and the control sub-systems fails guaranteeing the optimized control performance.

\subsection{Co-Design under Dynamics}

As opposed to separately designing the communication and the control sub-systems, the co-design of a wireless NCS offers a means to achieve the desired control performance.
Such a co-design presents an immediate challenge of increased computational complexity resulting from the expansion of parameter dimensions.
In essence, the communication and control co-design is a sequential decision-making process that adapts over time to the dynamics of a wireless NCS.
These dynamics not only emanate from the temporally and spatially fluctuating quality of communication and availability of radio resources, but are also intertwined with the operations of the control sub-system.
However, it is often not realistic to fully model the underlying dynamics of a real-world wireless NCS.
Additionally, deriving an accurate model that well approximates the intricate interactions between the communication and the control sub-systems remains impractical.

\section{Learning Framework Fortified with LLMs}
\label{framework}

In line with the previous discussions, this section aims to establish a learning framework for the communication and control co-design in the context of 6G networks.
The learning framework takes into account the sequential interactions between the communication and the control sub-systems over the time horizon.
Unlike the static environment assumption prevalent in most of the existing literature, we employ a discrete-time MDP to capture the dynamics inherent in a wireless NCS.

\subsection{Mathematical Formulation of an MDP}
\label{MDP_formulation}

\begin{figure}[t]
  \centering
  \includegraphics[width=21pc]{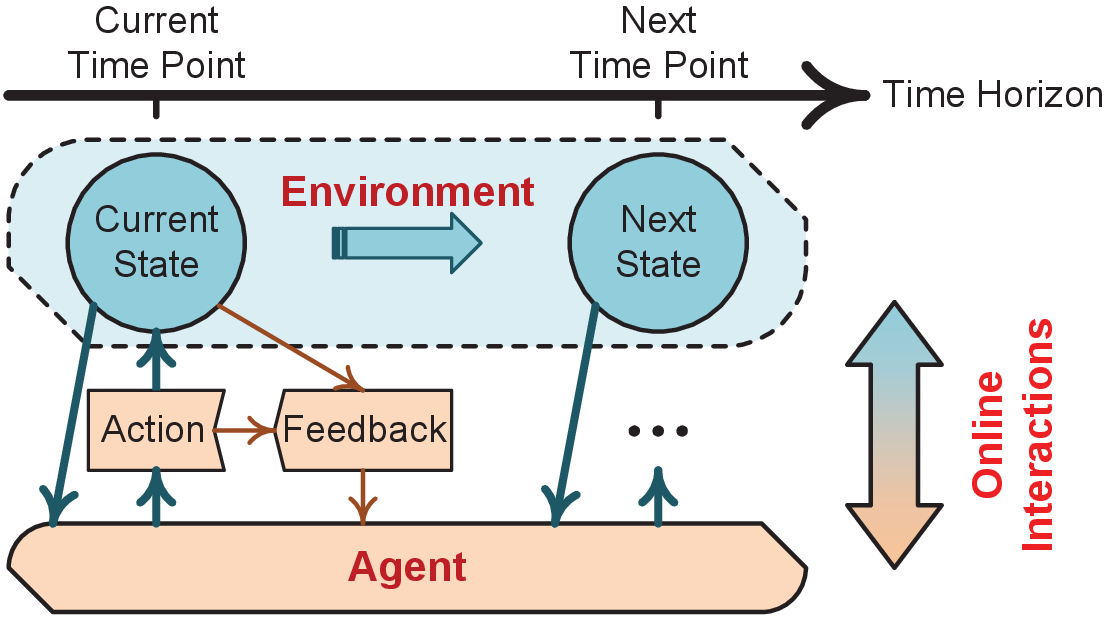}
  \caption{Diagram of a single-agent Markov decision process, during which an agent interacts with the surrounding environment over the discrete time horizon. Following a policy, the agent chooses an action based on the state observation at each current time point.}
  \label{MDP}
  \vspace{-0.5cm}
\end{figure}

In an MDP as depicted in Fig. \ref{MDP}, an agent (i.e., the decision-maker) engages in continuous interactions with the surrounding environment across the discrete time points.
The environment can be characterized by a finite set of states.
At each time point, the agent observes the current state, which is taken as the input to the decision-making policy that outputs an action from a set of feasible actions.
After executing the action, the surrounding environment transits from the current state to the next state, and meanwhile, the agent receives an instantaneous feedback.
Note that given an action under a state, the state transition adheres to the probability distribution over all possible subsequent states.
The feedback, which can be a reward or a cost signal, quantifies the direct impact of an action on a state.
Ultimately, the objective of the agent is to solve a decision-making policy that specifies the optimal actions under different states so as to optimize the expected long-term accumulated feedback.

When the MDP formulation is extended from single-agent to multi-agent settings, more than one agent simultaneously interact with the environment.
Consequently, both the controlled global state transition probability and the per-agent feedback become functions of the global state and the joint action, which are, respectively, composed of the local states and the local actions from all agents.
This simply indicates a coupling of policies, requiring that all agents share the local state information and respond to each other.

\subsection{Proposed Learning Framework}

The fundamental elements of a single-agent or multi-agent MDP formulation reveal the significant challenges in deriving an optimal decision-making policy, namely, the state representation from high-dimensional observations, the abstraction of action domains, the lack of a well-structured feedback model, and the need for an accurate controlled state transition probability function.
Fortunately, RL techniques learn the optimal policy without the controlled state transition probability.
However, in a practical wireless NCS, it remains daunting to employ RL to optimize the communication and control co-design due to the infinite problem size, the system intrinsic limitations as well as the rigorous control performance assessment.

\subsubsection{Integration of LLMs}

In this work, we propose a learning framework, which harnesses and integrates the potential of LLMs to enhance the remote control performance over the 6G wireless networks under dynamics.
LLMs, typically recognized as a variant of the Transformer neural network architecture, are characterized by billions of parameters and training on internet-scale data.
As the number of parameters and the volume of training data grow, LLMs tend to exhibit the capabilities of context sensitivity, logical reasoning, and nuance understanding \cite{Peng2024ICLR, Kwon2023ICLR}.
Specifically, an intelligent controller (i.e., the learning agent) fortified with an LLM is able to effectively fuse the radio frequency (RF) and non-RF modalities \cite{Nishio2019JSAC}, which interpret the real-world wireless NCS.
With the unified multi-modal data, the reasoning capability then facilitates the decision-makings (i.e., the actions in terms of communication and control co-design) through the LLM as the policy.
Additionally, the LLM allows the controller to strike a balance between instantaneous objectives, while being well-aligned with the optimized expected long-term control performance.

\begin{figure}[t]
  \centering
  \includegraphics[width=25pc]{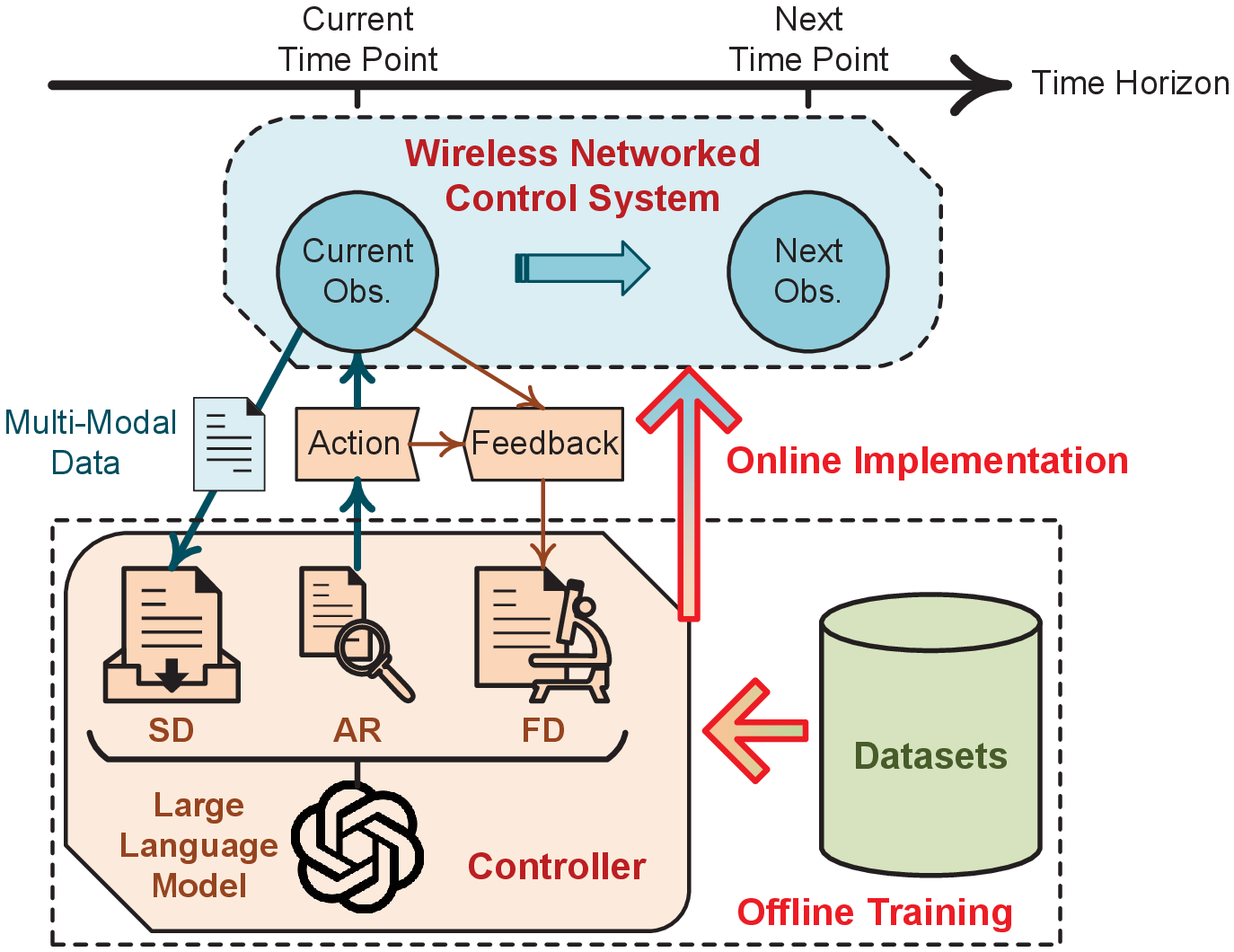}
  \caption{Offline reinforcement learning framework with large language models being integrated as different roles of state descriptor (SD), action recommender (AR) and feedback designer (FD).}
  \label{RL_LLMs}
  \vspace{-0.5cm}
\end{figure}

As shown in Fig. \ref{RL_LLMs}, we introduce in details the functioning roles of an LLM during the learning process for an optimized wireless NCS in 6G.
\begin{itemize}
    \item \textit{State Descriptor --}
    In a wireless environment, the observations made by the controller stem from multi-modal sources, involving the non-RF components, such as images and videos, in addition to the RF signals.
    LLMs can be employed as a state descriptor \cite{Peng2024ICLR}.
    The state descriptor is capable of extracting the spatial and temporal semantics from the environmental observations, thereby providing an accessible state representation as the input to the decision-making policy.

    \item \textit{Action Recommender --}
    Conventional RL techniques struggle with the low sampling efficiency, which becomes more apparent for the augmented problem of communication and control co-design under investigation.
    As an action recommender, LLMs make high-level recommendations to shrink the space of candidate actions \cite{Du2023ICML}.
    This enables the controller to pick a precise action directly based on the state information.

    \item \textit{Feedback Designer --}
    The difference in the feedbacks given different state-action pairs can be nuanced.
    However, the nuances turn to be noticeable solely from the perspective of communication or control.
    This asks for a feedback design that flexibly responds to the communication and control performance requirements in accordance with the system dynamics.
    Without a predefined model, we exploit LLMs as a feedback designer that autonomously generates reward or cost signals when the explicit control performance measures are impossible to specify \cite{Ma2024ICLR}.
\end{itemize}
We emphasize that the three distinct roles can be either jointly or separately integrated into our proposed learning, depending on the complexity of the communication and control co-design problem.
It is noteworthy that the decision-making policy employed by the controller can also be approximated through a sufficiently large neural network model, which addresses the state and action space explosions.

\subsubsection{Data-Driven Policy Optimization}

Nevertheless, the training of an optimal communication and control co-design policy can still be a significant challenge during the RL process in practice.
On one hand, the LLM-empowered controller collects experiences from repeatedly interacting with the wireless NCS to update the policy parameters.
Herein, an experience corresponds to a tuple of current state representation, action, immediate feedback and subsequent state representation.
On the other hand, the experience collection requires the real-time policy implementation.
Such parallel policy parameter updating and experience collection are compounded into the ``chicken and egg'' dilemma, which is central to the online RL.
However, the continuously collecting online experience data is time-consuming and costly.
Moreover, iterative policy parameter updating frequently generates random actions that pose risks to operating a real-world wireless NCS.
Despite being empowered by a pre-trained LLM, our proposed learning framework optimizes the communication and control co-design policy in an offline manner, leveraging a static dataset of a number of experiences (e.g., from the regular system operations).
Different from the online RL, the data-driven policy optimization during offline RL training relaxes the dependence on the interactions with the wireless NCS.

\subsubsection{Case Study on Age of Semantics-Aware Communication and Control Co-Design}

In a wireless NCS, such as smart manufacturing \cite{Noor22} and intelligent transportation \cite{JLin23}, to mention a few, it is crucial for the remote controller to keep up-to-date information of the physical process of interest, requiring the sensor to regularly update the status.
The concept of AoI, which measures the time since the most recent status update was received, has attracted extensive attention in the literature \cite{Yates21}.
However, AoI overlooks the semantics behind each status update.
This limitation motivates the proposal of age of semantics (AoS) to dig out the connection between the semantics at the sensor and the timely inference at the remote controller.
As a case study, we apply our proposed learning framework to the communication and control co-design with the purpose of optimizing AoS performance.

\noindent\textbf{System Model.} We concentrate on a discrete-time wireless NCS, where with the help of multiple half-duplex relays and an intelligent reflecting surface (IRS), the sensor updates the physical process of interest to the remote controller through the delivery of semantic samples.
The status of the process of interest is modelled using a finite-state Markov chain.
We assume that the interval $\tau$ between any two consecutive time points is constant.
At each time point, the remote controller decides whether the sensor should sample the process status.
If the process is sampled, the generated status update is compressed with the semantic extraction module.
Aided by the IRS, the sensor first transmits the semantic sample to the selected relay, and during the remaining time until the subsequent time point, the selected relay then forwards the received semantic sample to the remote controller in a decode-forward mode.
The remote controller is acknowledged if and only if the true process status aligns with the inference result of the status update, which is reconstructed from the semantic sample.
At the sensor, the energy is consumed by status sampling, semantic extraction and semantic sample transmission.
Considering the physical constraints, the transmit energy consumption can be calculated using the frequency bandwidth, the maximum transmit power, the channel gains to the selected relay and the IRS, as well as the data size of the semantic sample.
To quantify the semantics freshness at the remote controller, we define AoS as the time duration since the perfect inference of the process status.

\noindent\textbf{Problem Formulation and Solution.} Under the states $\{\mathbf{s}^t: t = 1, 2, \cdots\}$ across the discrete time points $t = 1, 2, \cdots$, the goal of the remote controller is to device a policy $\pi$ that optimizes the actions $\{\mathbf{a}^t = \pi(\mathbf{s}^t): t = 1, 2, \cdots\}$ such that the expected long-term reward $V(\mathbf{s}) = \mathsf{E}_{\pi}[\sum_{j = t}^\infty (\gamma)^{j - t} \cdot r(\mathbf{s}^t, \mathbf{a}^t) | \mathbf{s}^t = \mathbf{s}]$ is maximized.
Herein, the expectation $\mathsf{E}_{\pi}$ is calculated with respect to the probability distribution induced by $\pi$.
$\gamma \in [0, 1)$ denotes the discount factor.
At each time point $t$, the state $\mathbf{s}^t$ represents the information of AoS, channel gains of all the links and sensor-relay association state, the action $\mathbf{a}^t$ determines the status sampling and relay selection decisions, while $r(\mathbf{s}^t, \mathbf{a}^t)$ is the immediate reward signal of realized AoS at the remote controller and energy consumption at the sensor.
It is ready to see that the AoS-aware communication and control co-design falls into a single-agent MDP.
To solve the optimal policy, the conventional online RL requires the remote controller to explore any valid actions to access the wireless NCS in real-time, potentially increasing the collapse danger.

Though the assumptions made for simplification in this case study diminish the roles of LLMs, our proposed learning framework enables the remote controller to learn the optimal policy without the need of infinitely interacting with the system.
The major obstacle in data-driven policy learning originates from the overestimation owning to the extrapolation of out-of-distribution (OOD) actions \cite{Levine20}.
One direction is that the remote controller estimates the state-action function defined by $Q(\mathbf{s}, \mathbf{a}) = \mathsf{E}_{\pi}[\sum_{j = t}^\infty (\gamma)^{j - t} \cdot r(\mathbf{s}^t, \mathbf{a}^t) | \mathbf{s}^t = \mathbf{s}, \mathbf{a}^t = \mathbf{a}]$, $\forall (\mathbf{s}, \mathbf{a})$ through minimizing the loss function as in \cite{Fujimoto19} and  constraining the feasible action space to support the static dataset.
However, the estimated policy is highly sensitive to the dataset quality since the remote controller has no opportunity to correct the actions with new experiences.
In order to ease the extrapolation of OOD actions, our learning framework constructs an augmented loss function to peculiarly penalize the estimated state-action function, with which the estimated policy improves upon the empirical policy and the value of an OOD action is larger than that of an action from the dataset by a certain margin.
Note that the empirical policy is given as the frequency of an action appearing in the dataset with the same current state.

\noindent\textbf{Results and Analysis.}
To evaluate the performance from our proposed learning framework, we set up a wireless NCS, where there are five relays and the IRS is deployed to have line-of-sight links with the sensor, the relays and the remote controller.
The interval duration is set to be $\tau = 0.1$ seconds.
At each time point, the channel gains of all links are independent and identically distributed \cite{Bjor20}, while the process of interest is in one of $9$ states.
We assume that the process of interest keeps staying in the same state during the next time point with a probability of $\alpha$ and the probability of switching to another different state is $(1 - \alpha) / 8$.
At the remote controller, we design a neural network with one hidden layer containing $64$ neurons and use $\beta$ to denote the probability of accurate inference of the reconstructed status update.
The data size of a semantic sample is assumed to be $6.2 \cdot 10^6$ bits.
The system frequency bandwidth is $10$ MHz, while the transmit powers at the sensor and the relays are $1$ Watts.
For the comparison purposes, we select three baselines: the advantage actor-critic (A2C) scheme \cite{Mnih16}, the Random scheme and the conservative Q-learning (CQL) scheme \cite{Kumar20}.
With the Random scheme, the remote controller employs a policy that applies a uniform probability distribution over the feasible actions at each time point.
For all experiments, we collect two datasets, namely, Expert Data from the A2C scheme after convergence and Random Data from the Random scheme.

\begin{figure}[t]
  \centering
  \includegraphics[width=25pc]{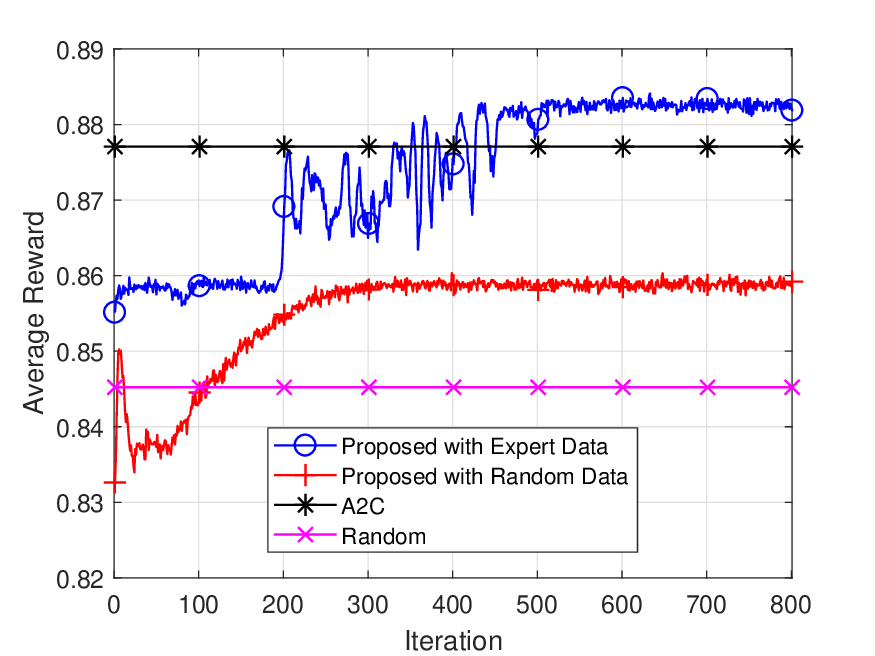}
  \caption{Convergence performance of our proposed learning framework, where $\alpha = \beta = 0.5$ and the IRS is with $75$ reflecting elements. Each point on the curves of our proposed learning framework corresponds to an average value of $10^5$ reward realizations from testing the policy that is trained offline at each iteration.}
  \label{convergence}
  \vspace{-0.5cm}
\end{figure}

\begin{figure}[t]
  \centering
  \includegraphics[width=25pc]{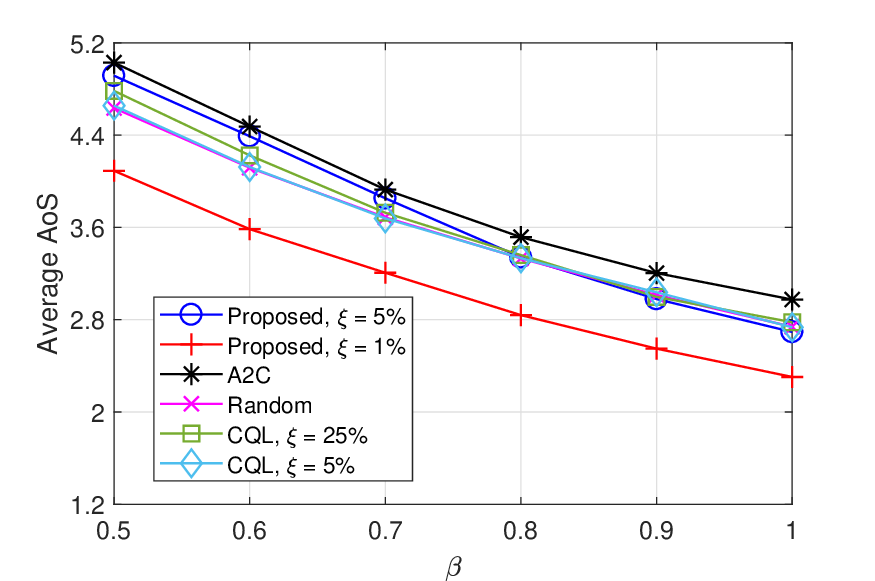}
  \caption{Average AoS performance versus accurate inference probability, where $\alpha = 0.3$ and the IRS is with $25$ reflecting elements.}
  \label{simu2_1}
  \vspace{-0.5cm}
\end{figure}

\begin{figure}[t]
  \centering
  \includegraphics[width=25pc]{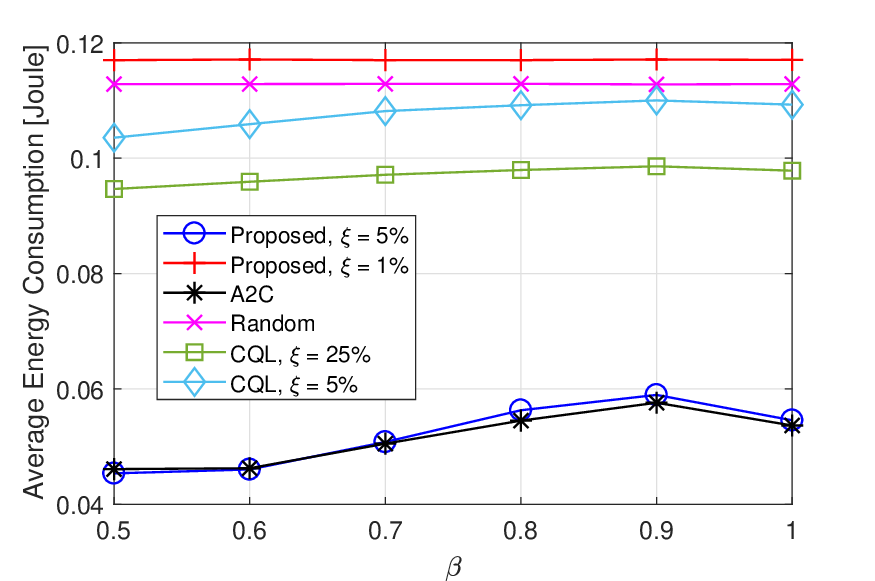}
  \caption{Average energy consumption versus accurate inference probability, where $\alpha = 0.3$ and the IRS is with $25$ reflecting elements.}
  \label{simu2_2}
  \vspace{-0.5cm}
\end{figure}

\begin{figure}[t]
  \centering
  \includegraphics[width=25pc]{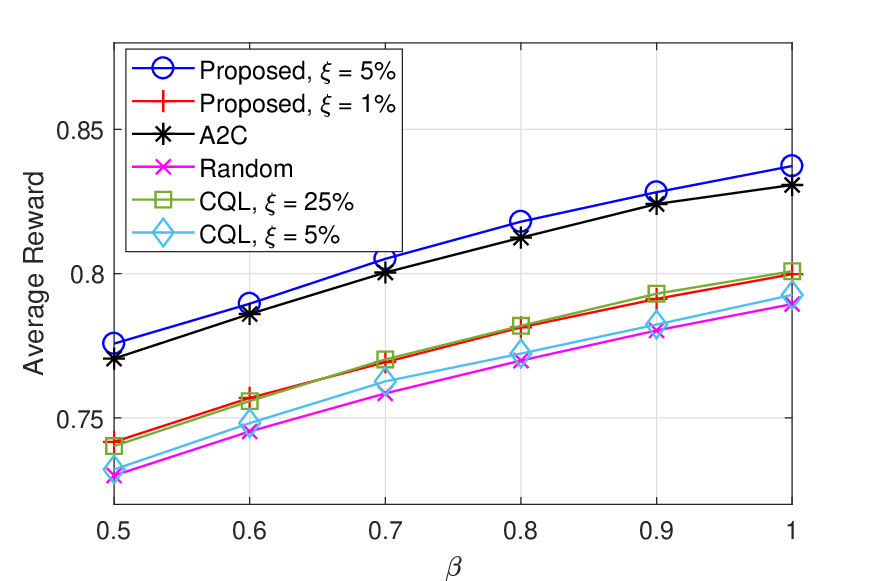}
  \caption{Average reward performance versus accurate inference probability, where $\alpha = 0.3$ and the IRS is with $25$ reflecting elements.}
  \label{simu2_3}
  \vspace{-0.5cm}
\end{figure}

We begin by assessing the convergence speed during the training of our learning framework as in Fig. \ref{convergence}.
The curves also show the average reward performance from testing the trained A2C and the Random schemes.
Notably, our learning framework, trained with Expert Data and Random Data, outperforms the A2C scheme and the Random scheme, respectively, after converging within $600$ iterations.
This improvement can be attributed to the delicate trade-off between exploration and exploitation for any online RL schemes.
To examine the robustness of our learning framework to the quality of dataset, we mix Expert Data and Random Data to get new training datasets and let $\xi$ denote the fraction of Expert Data.
By varying the value of $\beta$, Figs. \ref{simu2_1}, \ref{simu2_2} and \ref{simu2_3} illustrates the obtained average AoS, average energy consumption and average reward performance.
As the value of $\beta$ increases, the remote controller keeps fresher inference results about the process of interest, which in turn demands the sensor to sample the process status more frequently, verifying a decrease in average AoS as in Fig. \ref{simu2_1} and an increase in average energy consumption as in Fig. \ref{simu2_2}.
When the value of $\beta$ is sufficiently large, the sensor no longer needs to retain a high sampling frequency, verifying a decrease in average energy consumption as in Fig. \ref{simu2_2}.
With a small enough fraction of Expert Data (e.g., $\xi = 1\%$), our learning framework struggles to distinguish a new dataset from Random Data.
When trained using datasets containing only $\xi = 5\%$ Expert Data, our learning framework achieves moderately better average reward performance than the A2C scheme, but significantly outperforms the CQL scheme trained using datasets with $\xi = 25\%$ Expert Data, as in Fig. \ref{simu2_3}.

\section{Open Issues and Future Research Directions}
\label{directions}

In the preceding section, we formulate the problem of communication and control co-design in 6G as an MDP, and propose an offline RL based solution framework that is fortified by LLMs.
Our learning framework explores the promising potentials of integrating an LLM as the state descriptor, the action recommender and the feedback designer.
In what follows, we delve into the open issues and future research directions for implementing our learning framework, spanning from modelling and theoretical considerations to practical realities.
\begin{itemize}
    \item \textit{Multi-Modal Feature Abstraction --}
    Due to the increasingly complex infrastructure deployments in the 6G era, we lack a deep understanding of the wireless NCS, which is challenging to be exactly described using the existing statistical models.
    Spurred by the advances in machine learning and computer vision, both RF and non-RF modalities can be exploited to enrich the state observations.
    However, efficiently fusing multiple data modalities to extract the semantics is critical for system feature abstraction.
    Through the integration of an LLM, the controller is capable of retrieving the accessible state representations.
    Nevertheless, open issues remain in compression of LLMs for practical implementations on resource-constrained nodes and generalization of LLMs to heterogeneous data modalities.

    \item \textit{Communication and Control Interactions --}
    The selection of an action at each time point influences both the immediate and the future control performance under the MDP formulation of communication and control co-design.
    To strike an interactive balance between the communication and control sub-systems, LLMs leverage the state information and act as a prior to propose a high-level compact action space.
    As such, the controller proceeds to bottom out the low-level actions, which accelerates the offline training processing of our learning framework.
    It remains a bottleneck how to identify the space of actions that is highly relevant to the one-shot and long-term objectives conditioned on the state information at any time point.

    \item \textit{Feedback Hacking Risks --}
    Due to the communication and control interactions, the control performance is multi-attributed and each of the attributes contributes to the feedback (i.e., the reward signal or the cost signal).
    These attributes include the measurements of network-induced latency, data loss, quantization errors in signal sampling, and various effects from the system dynamics.
    Yet there does not exist a rigorous method for the definition of a per-time point feedback function that quantitatively specifies and combines all such attributes into a single real-valued number.
    Even with the feedback function trained by an LLM, it is still possible to extrapolate the trajectory of undesirable actions that optimize merely the accumulated feedback but not the control performance, due to the feedback hacking \cite{Li19}.

    \item \textit{Non-Stationary Generalization --}
    During the offline training of our learning framework, the policy is estimated upon the basis of the empirical policy, which is calculated from the static dataset under the assumption of a stationary wireless NCS.
    When the statistics of system dynamics changes (i.e., in a non-stationary wireless NCS), the control performance resulting from implementing the estimated policy cannot be guaranteed.
    Without the loss of generality, a non-stationary wireless NCS can be mathematically represented through a set of MDPs with independent state transition probabilities.
    Meta learning assists in finding a meta policy that quickly adapts to a new MDP \cite{Finn17}.
    To enhance the control performance, the meta policy needs to be fine-tuned with the newly collected interaction experiences.
    It is important to notice that during the fine-tuning, the meta policy avoids random actions.

    \item \textit{Data Security and Privacy in Implementation --}
    An LLM, which takes any modality data as the input, plays a crucial role in our proposed learning framework, giving rise to the security and privacy concerns related to data collection/transmission, control signalling, and so forth.
    Due to the openness of a wireless communication network, the implementation of our learning framework is vulnerable to the adversarial attacks.
    How to detect and counteract the attacks for the protection of a wireless NCS in 6G is a direction worth exploring.
    In addition, the training of our learning framework relies on massive datasets.
    To prevent the severe and unintended consequences from harmful data, it is imperative to deploy robust data validation techniques to remove the sensitive information.

\end{itemize}

\section{Conclusions}
\label{conclusions}

In this article, we concentrate our focus on the investigation of communication and control co-design over a 6G wireless network.
The dynamics inherent in a wireless NCS necessitates the adoption of an MDP to mathematically formulate the co-design problem.
To approach the optimal policy, we propose an offline learning framework.
Our proposed learning framework leverages LLMs in the RL process to enable the controller to optimally configure the communication and control parameters over the discrete time horizon.
A case study on AoS-aware communication and control co-design evaluates the performance gains from our learning framework.
Last but not least, we identify the major open issues and research opportunities associated with the implementation of our proposed learning framework.




\ifCLASSOPTIONcaptionsoff
  \newpage
\fi


\end{document}